\documentclass[reviewcopy]{elsart}
\usepackage{epsfig,multicol,graphics}
\newcommand{\kalium}{{$^{40}$K}}
\newcommand{\radon}{{$^{222}$Rn}}
\newcommand{\radium}{{$^{226}$Ra}}
\newcommand{\polonium}{{$^{214}$Po}}
\newcommand{\bismuth}{{$^{214}$Bi}}
\newcommand{\carbon}{{$^{14}$C}}
\newcommand{\uran}{{$^{238}$U}}
\newcommand{\thorium}{{$^{232}$Th}}
\usepackage{amssymb}

\begin{document}

\begin{frontmatter}



\title {Phenylxylylethane (PXE): a high-density, 
high-flashpoint organic liquid scintillator for
applications in low-energy particle and astrophysics experiments}

\begin{center}
{Borexino Collaboration}
\end{center}
\author{}{
{H.O.~Back$^{a}$},
{M.~Balata$^{b}$},
{A.~de~Bari$^{c}$},
{T.~Beau$^{d}$},
{A.~de~Bellefon$^{d}$},
{G.~Bellini$^{e}$},
{J.~Benziger$^{f}$},
{S.~Bonetti$^{e}$},
{A.~Brigatti$^{e}$},
{C.~Buck$^{g}$
% ...
}
}

\address{
{$^{a}${Physics Department, Virginia Polytechnic Institute and State University,
Robeson Hall, Blacksburg, VA 24061-0435, USA}}\\ 
{$^{b}${I.N.F.N Laboratori Nazionali del Gran Sasso, SS 17 bis Km 18+910, I-67010 Assergi(AQ), Italy}}\\
{$^{c}${Dipartimento di Fisica Nucleare e Teorica Universit\`a and I.N.F.N., Pavia, Via A. Bassi, 6 I-27100,
Pavia, Italy}}\\ 
{$^{d}${Astroparticule et Cosmologie, 10, rue Alice Domon et Leonie Duquet, F-75025 Paris cedex 13}}\\ 
{$^{e}${Dipartimento di Fisica Universit\`a and I.N.F.N., Milano,
Via Celoria, 16 I-20133 Milano, Italy}}\\ 
{$^{f}${Dept. of Physics,Princeton University, Jadwin
Hall, Washington Rd, Princeton NJ 08544-0708, USA}}\\ 
{$^{g}${Max-Planck-Institut f\"ur Kernphysik,Postfach 103 980, D-69029 Heidelberg, Germany}}\\ 
{$^{h}${Dept. of Physics, Queen's University Stirling Hall, Kingston, Ontario K7L 3N6, Canada}}\\ 
{$^{i}${Technische Universit\"at M\"unchen, Physik Department E15, James Franck Stra\ss e, D-85747, Garching, 
Germany}}\\ 
{$^{j}${Joint Institute for Nuclear Research, 141980 Dubna, Russia}}\\ 
{$^{k}${Dept. of Physics Massachusetts Institute of Technology,Cambridge, MA 02139, USA}}\\ 
{$^{l}${Dipartimento di Chimica Universit\`a, Perugia, Via Elce di Sotto 8, I-06123 Perugia, Italy}}\\ 
{$^{m}${RRC Kurchatov Institute, Kurchatov Sq.1, 123182 Moscow, Russia}}\\ 
{$^{n}${Dipartimento di Fisica Universit\`a and I.N.F.N., Genova, Via Dodecaneso,33 I-16146 Genova, Italy}}\\ 
{$^{o}${KFKI-RMKI, Konkoly Thege ut 29-33 H-1121 Budapest, Hungary}}\\
{$^{p}${Bell Laboratories, Lucent Technologies, Murray Hill, NJ 07974-2070, USA}}\\ 
{$^{q}${M. Smoluchowski Institute of Physics, Jagellonian University, PL-30059
Krakow, Poland}}}

\thanks{
{{Now at Massachusetts Institute of Technology, NW17-161, 175 Albany St.
Cambridge, MA 02139} }\\
{$^{2}$~{Now at Technische Universit\"at M\"unchen, James Franck Stra\ss e, D-85747 Garching, Germany}}\\ 
{$^{3}$~{Now at Sudbury Neutrino Observatory, INCO Creighton Mine, P.O.Box 159
Lively, Ontario, Canada, P3Y 1M3}}\\
{$^{4}$~{Now at Universit\"at Hamburg, Luruper Chaussee 149, D-22761 Hamburg,
Germany} }\\
{$^{5}$~{Now at Dept. of Physics, Queen's University Stirling Hall, Kingston, Ontario K7L 3N6, Canada}}\\ 
{$^{6}$~{Now at Max-Planck-Institut fuer Kernphysik, Heidelberg, Germany}}\\
{$^{7}$~{Now at Max-Planck-Institut fuer Plasmaphysik, Boltzmannstr.2 D-85748
Garching, Germany}}\\
{$^{8}$~{Now at Center for Cosmological Physics, University of Chicago, 933 E.56$^{th}$St., Chicago, IL 60637}}\\
{$^{9}$~Marie Curie fellowship at LNGS}\\ 
{$^{10}$~On leave of absence from St. Petersburg Nuclear Physics Inst. - Gatchina, Russia} \\ 
{$^{11}$~On leave of absence from Institute for Nuclear Research, MSP 03680, Kiev, Ukraine} \\ 
{$^{12}$~On leave of absence from KFKI-RMKI, Konkoly Thege ut 29-33 H-1121
Budapest, Hungary} \\
{$^{13}$~Now at Lockhead Martin Corporation, Sunnyvale CA} \\
{$^{14}$~Now at Physics Department, Virginia Polytechnic Institute and 
State University, Robeson Hall, Blacksburg, VA 24061-0435, USA} \\
{$^{15}$~Deceased} \\
{$^{16}$~\underline{Corresponding authors:}\\
Stefan Sch\"onert, email: stefan.schoenert@mpi-hd.mpg.de,\\
Marianne G\"oger-Neff, email: marianne.goeger@ph.tum.de.}\\
}

\begin{abstract}
We report on the study of a new liquid scintillator target 
for neutrino interactions in the framework of the research
and development program of the Borexino solar neutrino experiment.
The scintillator consists of  
1,2--dimethyl--4--(1--phenylethyl)--benzene  
(phenyl--o--xylylethane, PXE) as solvent and 
1,4-diphenylbenzene (para-Terphenyl, p-Tp) as primary 
and 1,4-bis(2-methylstyryl)benzene (bis-MSB) as secondary solute. 
The density close to that of water and the high flash point makes 
it an attractive
option for large scintillation detectors in general. The study 
focused on optical properties, radioactive trace impurities 
and novel purification 
techniques of the scintillator. Attenuation lengths of the 
scintillator mixture of 12~m at 430~nm 
were achieved after purification with an alumina column. 
A radio carbon isotopic ratio of  
$\rm ^{14}C/^{12}C = 9.1 \times 10^{-18}$ has been measured in the 
scintillator. Initial trace impurities, 
e.g.  \uran\ at $3.2\times 10^{-14}$~g/g 
could be purified to levels  below $1\times 10^{-17}$~g/g by silica gel 
solid column purification.
\end{abstract}

\end{frontmatter}

\section{Introduction}
Organic liquid scintillators are used in large quantities for rare event 
detection in particle astrophysics. 
The main objective in these experiments is the real time 
spectroscopy of neutrinos from steady-state sources such as 
the Sun, nuclear reactors and from beta decays in the crust and 
mantle of the Earth, as well as from transient sources such as 
supernovae. 

Despite the large target mass of several hundreds of tons, 
the signal rates of the steady-state sources are typically in 
the range of a few events per day down to a few events per year
at MeV or sub-MeV energies. Thus, background signal rates created
by radioactivity and cosmic ray interactions need to be extremely low. 

Low backgrounds can be achieved by locating the 
detectors deep underground to suppress the cosmic ray muon flux, 
 shielding the scintillator target against the ambient 
radioactivity from the surrounding rocks, and suppressing and 
removing radioactive impurities present in trace amounts in the detector 
and ancillary systems as well as in the liquid scintillator itself.
This concept of background reduction has been pioneered by the 
Borexino collaboration \cite{BX-ST}
in the Counting Test
Facility (CTF) \cite{CTF-NIM,CTF-C14,CTF-AP} and is implemented in the 
Borexino detector, and similarly, in the 
KamLAND experiment \cite{KL}.

This paper summarizes the study of  
1,2--dimethyl--4--(1--phenylethyl)--benzene  
(phenyl--o--xylylethane, PXE), 
a new scintillator solvent the key characteristics of 
which are its high density  ($0.988\,{\rm g/cm}^3$) and high flash 
point (145$^o$C). This scintillator solvent has been investigated  
as a `back-up solution' for the Borexino experiment. 
The design of choice is based on  
1,2,4-trimethyl\-benzene (pseudocumene, PC), both as buffer liquid and 
as neutrino target. 
  
Since the density of PXE is close to that of water, even large 
scintillator targets can be submerged in water - serving as 
a shield against ambient radiation - while 
creating only modest buoyancy forces on the scintillator 
containment vessel. 
Moreover, a detector with  water as shield and PXE as target material  
provides a substantial higher fiducial target mass because of the 
improved shielding performance against external radiation
compared to a detector with identical dimensions, but  
with both shield and target of organic liquids with standard
densities ($\sim 0.9$~g/cm$^3$). 
Finally, a PXE-water configuration reduces the overall inventory 
of organic liquid in the detector systems, e.g. to about one fourth 
in the case of Borexino. This,
depending on  national regulations,  may have an impact on the legal 
classification of the detector systems
and therefore on the safety and operational aspects of the 
experiment. A further asset of PXE is its high flash point 
simplifying safety systems relevant for transportation, 
handling and storage. According to regulations by the United Nations (UN),
PXE is legally non-hazardous for transportation purposes 
and no special United Nation number code applies \cite{CFR}.
In a paper by Majewski et al. \cite{Majewski} PXE
has been described as a relatively safe solvent with very low toxicity
compared to standard liquid scintillators. 
The Double Chooz reactor neutrino experiment \cite{doublechooz}, aiming at a measurement 
of the mixing angle $\theta_{13}$, will use as target a scintillator mixture based on PXE.

The study reported here of PXE  as a solvent for a low-background
scintillator for solar neutrino spectroscopy was carried out within 
the Borexino project. Research on PXE scintillators started
in 1995 with laboratory measurements focusing on  
optical properties and radio purification techniques with 
solid columns. After completion of the laboratory scale study,  
about five tons of PXE solvent for testing on prototype scale 
with the Counting Test Facility (CTF) of Borexino were 
acquired in 1996.
Fluors were added on site in hall C of the Laboratori Nazionali del 
Gran Sasso (LNGS)
and the final scintillator was purified with 
Module-0, a solid column purification and liquid handling system 
\cite{Mod0}. The CTF was loaded with PXE scintillator in October 1996
and first data were acquired until January 1997.
The quality of data was limited because only a small fraction 
of the photomultipliers in the CTF were operational. 
The PXE scintillator was unloaded from the CTF after 
the shut down of the detector 
in July 1997, and moved back into the storage tanks
of Module-0. Further batch purification operations were carried out
during the period October until December 1997.
Samples for neutron activation analysis were taken 
to monitor the achieved radio purity after 
each operation. After reconstruction of the CTF during 1999, the 
PXE scintillator was reloaded into the CTF and measured 
from June to September 2000. The main objective during this 
period was the analysis of radio purity and optical
properties in a large volume detector. 
Beyond this scope,
physics limits on electron instability \cite{el-decay},
nucleon instability \cite{nucl-decay}, 
neutrino magnetic properties \cite{nu-elm}, and on 
violations of the Pauli exclusion principle \cite{pauli-excl}
could be derived from measurements with the PXE scintillator.

The paper is structured as follows: Section~2 summarizes the physical and 
chemical properties of the PXE solvent. Section~3 describes the optical 
properties of the solvent as well as the mixed scintillator. Section~4 
is dedicated to the large scale test of the PXE scintillator with the 
CTF including the scintillator preparation, purification and analysis
of trace impurities, and conclusions are given in Section~5.

\section{Physical and chemical properties of PXE}

PXE is a clear, colorless liquid with an aromatic odor. 
It is an industrial product with different applications, as for example: 
insulating oil in high voltage transformers and capacitors and  
as oil for pressure sensitive paper. 
PXE is produced  by reacting styrene and xylene 
using an acidic catalyst.  It is then washed with 
water and distilled to improve the purity.
Its final industrial purification step uses a solid column.

\begin{figure}
\epsfig{width=8cm,file=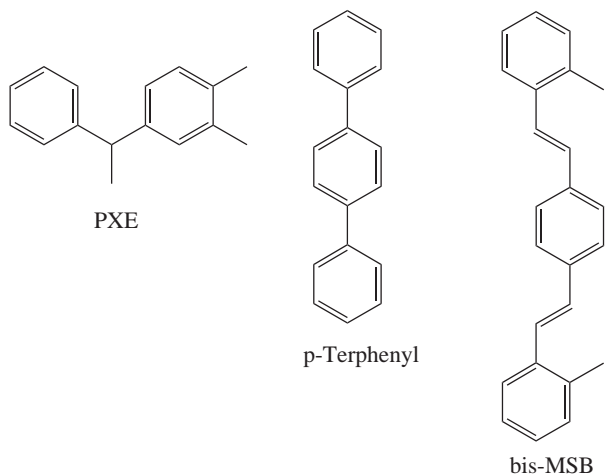}
\caption{\label{f:pxemol}Molecular structure of 1,2-dimethyl-4-(1-phenylethyl)-benzene   
(phenyl-o-xylylethane, PXE),  1,4-diphenylbenzene (para-Terphenyl, p-Tp) 
and 1,4-bis(2-methylstyryl)benzene (bis-MSB).}
\end{figure}

PXE has the molecular formula 
$\rm C_{16} H_{18}$ with a weight of 210.2 g/mol.
Its molecular structure is shown in 
Figure\,\ref{f:pxemol}.

Chemical and physical details which are of relevance for detector design,
safety and operational aspects 
are listed in Table\,\ref{t:pxe-chem-phys}. 
The key features are its
high density of 0.988~g/cm$^3$, its low vapor pressure and 
high flash point of  145~$^o$C. It is therefore classified as 
a non-hazardous liquid.

For our test, 
the standard production scheme at Koch Chemical 
Company, Corpus Christi, Texas, USA\footnote{Koch Special Chemical 
Company stopped production of PXE in 2002. 
Nippon Petrochemicals Co., Ltd., Japan, produces PXE as an equal 
mixture of ortho, para and meta isomers}
was modified omitting  
the final column purification 
at the company, since it was expected that the 
clay column material might leach off radio active impurities, such as
uranium and thorium.
Instead a column purification system was built
and operated using silica gel in the underground laboratories 
of the LNGS. Details are discussed below.


\section{Optical properties}
The optical and scintillation properties of the pure PXE solvent, 
of selected fluors, and of PXE-fluor  mixtures 
have been investigated by UV/Vis spectrometry, by fluori\-metry and by
excitation with ionizing radiation. The objective was to optimize
the scintillator performance by maximizing its light yield and 
attenuation length and minimizing its scintillation decay time. 
Methods to remove optical impurities 
were studied, since impurities can potentially reduce light yield and 
attenuation length. All optical properties presented in this section
are derived from laboratory size samples  up to a few hundred ml's. 
Attenuation length measurements typically are done in `one-dimension' only. 
Scattered light, elastic or inelastic, is undetectable in these measurements 
whereas in large volume applications with 4$\pi$ geometry, the scattered photons
are not necessarily lost. 
The performance of the PXE scintillator 
in a large volume detector, taking scattering into account,  
has been studied in CTF and is discussed in Sec.~4.4.

\subsection{Solvent properties}
\begin{figure}
\epsfig{width=12cm,file=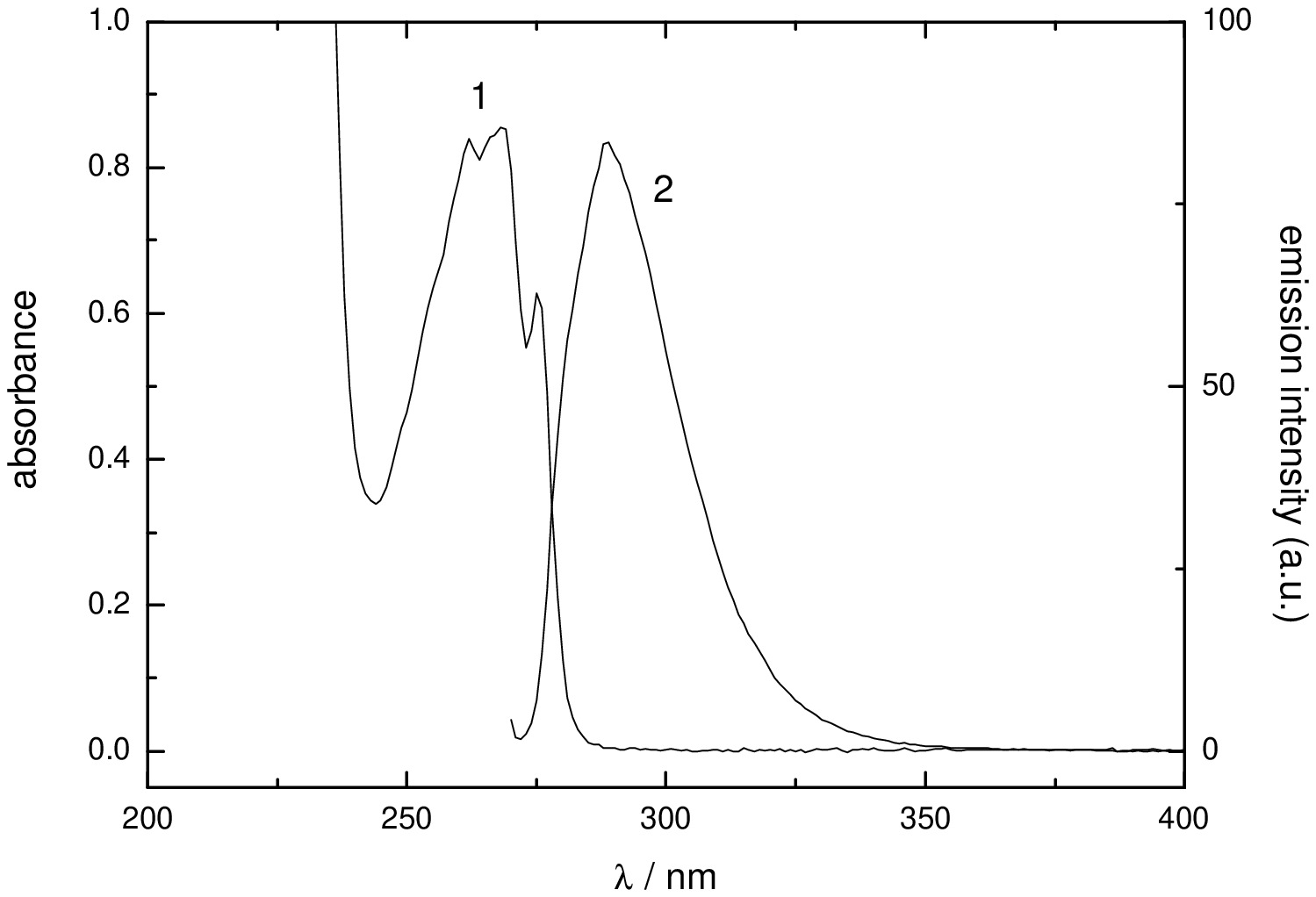}
\caption{\label{f:pxe-emi-abs}Absorption (1) and emission (2) spectrum of PXE 
diluted in cyclohexane.}
\end{figure}

PXE diluted in cyclohexane shows an absorption maximum at 267\,nm. The 
emission spectrum after excitation at this wavelength peaks at about 
290\,nm as displayed in Figure~\ref{f:pxe-emi-abs}. 
The fluorescence lifetime after excitation at $\lambda_{exc}=267\,{\rm nm}$
measured in diluted solutions of cyclohexane
shows an exponential decay with a lifetime $\tau$ of $\sim 22$\,ns. 
Preliminary samples obtained from 
Koch had various optical impurities with absorption 
bands around 300, 325, 360 and 380\,nm. They could be observed 
both by UV/Vis and 
fluorimetric measurements. Passing the solvent through a column 
with acidic alumina reduced the absorption peaks.  
The band at 380\,nm was reduced most efficiently.

The PXE used  for the 5~t test, as described Section 4, 
still had various optical impurities, but at much
reduced levels. The attenuation length at 430\,nm ($\Lambda_{430}$), 
defined as $I(x) =I_0 \cdot \exp(-x/\Lambda_{430})$, with the initial 
intensity $I_0$ and attenuated intensity $I(x)$
after the optical path length $x$, was  
$\Lambda_{430} = 10.2 \,{\rm m}$ in the pure PXE solvent. Attenuation lengths
were measured with a Varian Cary 400 UV/Vis photospectrometer in 10\,cm cuvettes.

\subsection{Scintillator properties}\label{scintprop}

\begin{figure}
\epsfig{width=8cm,file=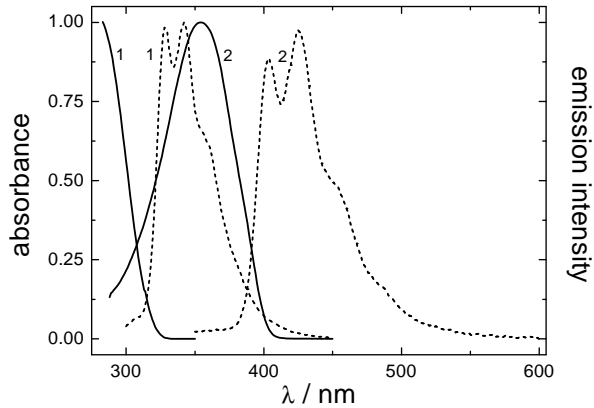}
\caption{\label{f:pxe-fluors}Absorption (full lines) and emission (dashed lines) 
spectra of the primary fluor p-Tp (1)
and the secondary fluor bis-MSB (2).}
\end{figure}

A tertiary scintillator system was chosen in order to shift the emission
wavelength to about 430 nm, well above the absorption bands of 
the residual optical impurities. This is achieved by using 
1,4-diphenylbenzene (para-Terphenyl, p-Tp) as primary solute
and 1,4-bis(2-methylstyryl)benzene (bis-MSB) as secondary solute.
Absorption and emission spectra of PXE in cyclohexane and those of 
the fluors in PXE are shown in 
Figures~\ref{f:pxe-emi-abs} and \ref{f:pxe-fluors}.
The latter shows that the emission spectrum of p-Tp is 
satisfactorily matched by the absorption spectrum of the wavelength shifter 
bis-MSB, and therefore an 
efficient energy transfer is expected in this solvent.

The scintillator properties  were tested with
p-Tp concentrations at 2 and 3~g/l (close to the 
solubility limit) and with bis-MSB at 20~mg/l. 
A scintillation yield of 88~(93)\,\% for PXE with an addition of 
p-Tp (2.0 (3.0)~g/l)/bis-MSB(20~mg/l) 
with respect to a scintillator based on 
1,2,4-trimethylbenzene (PC) and 1.5~g/l of 2,5-diphenyloxazole (PPO)
has been found (uncorrected for the PMT sensitivity).
Attenuation lengths of $\Lambda_{430}=2.6 $~m to 3.2~m 
of the scintillator mixture have been measured
depending on the sample treatment (cf. Section 4.2).

\begin{figure}
\epsfig{width=12cm,file=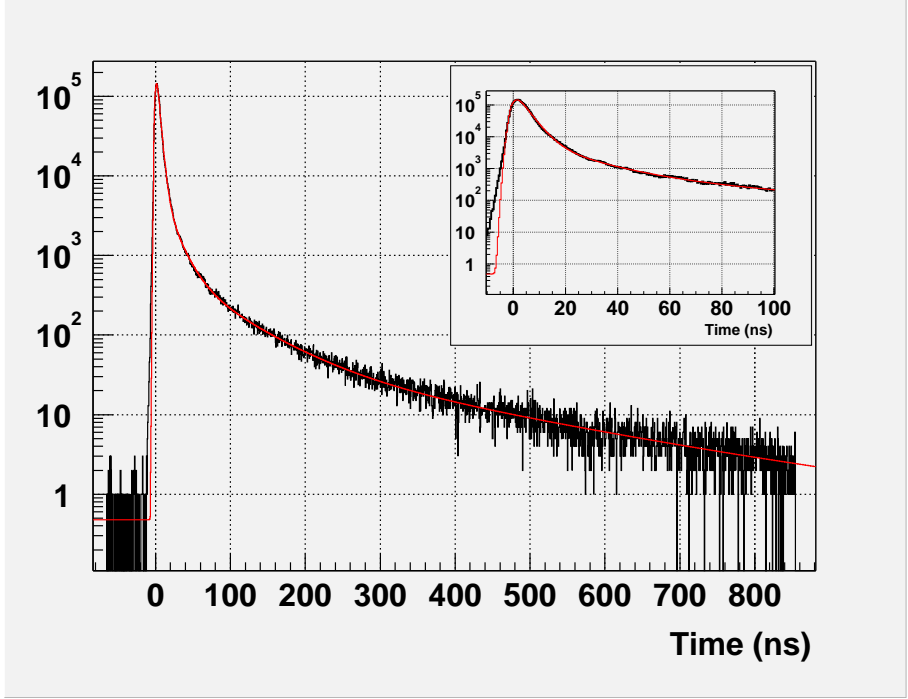}
\caption{\label{fpxecs137a}Photon emission time distribution of 
PXE/(2.0 g/l)p-Tp/(20 mg/l)bis-MSB 
scintillator after excitation with gamma radiation from $^{137}$Cs decay. The 
continuous curve shows the fit to the data with a four-exponential decay model.
The insert shows enlarged the first 100 ns.}
\end{figure}

The fluorescence decay time
of the p-Tp(2.0~g/l)/bis-MSB(20~mg/l) 
scintillator mixture measured by fluorimetry 
after excitation at 267 nm  
shows a fast component $\tau $ of 3.7~ns. At 3~g/l p-Tp, 
the decay constant is shortened to 3.2~ns. The photon emission 
probability density function (pdf) was further studied with ionizing radiation
in order to investigate the contribution of long lived 
triplet states and the possibility to 
use the emission time for discrimination of alpha versus beta particles
(pulse shape discrimination). 
The response to gamma radiation has been measured for a 
concentration of 2.0~g/l and 3.0~g/l p-Tp with a $^{137}$Cs source, 
and to alpha radiation at a concentration of 3.0~g/l using a 
$^{210}$Po alpha source \cite{tesilombardi,motta}.
The pdf can be parametrized
by the weighted sum of four exponentials 
$\sum_i{(q_i/\tau_i) \exp(-t/\tau_i)}$ with 
the parameters given in Table
\ref{t:em}. The emission time distribution is shown in fig.~\ref{fpxecs137a}.
In addition to an increased slow component which can be used for pulse shape 
discrimination, alpha particles emit less light because of the high 
ionization density compared to electron
or gamma radiation. The quenching factor is mainly solvent dependent and
has been measured  for different alpha decays in the $^{238}$U chain, using 
\radon\ loaded scintillator.  
The results are given in Table\,\ref{t:quench}
~\cite{neff-da}.

\begin{figure}
\epsfig{width=10cm,file=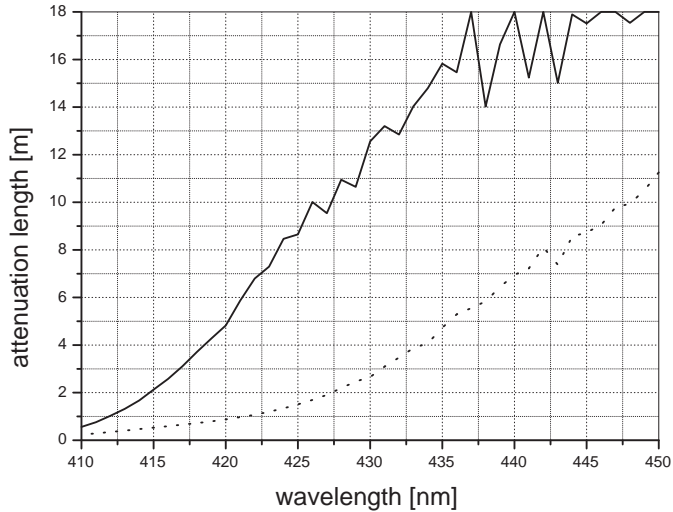}
\caption{\label{fpxeatt}Attenuation length of the PXE/p-Tp/bis-MSB scintillator before 
(dashed line) and after (solid line) purification with an alumina column.}
\end{figure}

Light attenuation of the standard mixture  
(PXE/(2.0 g/l)p-Tp/(20 mg/l)bis-MSB) which has been used
in the CTF (cf. Sec. 4) has been measured at 
various steps during the preparation of the scintillator in 1996 
and after filling the CTF. 
After several purification steps of the mixed scintillator 
through a silica gel column in Module-0 in order to remove radio-impurities
(cf. sections below), 
attenuation lengths in the range between 2.6 and 3.0~m at 430~nm were measured.
After completion of the PXE measurements with CTF and subsequent batch 
purification
operations, the scintillator has been stored
in barrels under nitrogen atmosphere. In 2003, this  scintillator was 
used in the frame of the LENS project~\cite{llbf} and optical properties 
were remeasured to check for degradation with time. 
The attenuation length as well as the light yield were 
unchanged with respect to the measurements in 1996. 
Passing this scintillator mixture through a weak acidic alumina column
increased the attenuation lengths to 12~m as displayed in 
Figure~\ref{fpxeatt}~ 
while retaining the scintillation yield \cite{buck}.  

\section{Large scale test of PXE with the Borexino Counting Test Facility}
The main objectives of the large scale test in the 
Borexino prototype detector
(Counting Test Facility, CTF \cite{CTF-NIM})
were the study of 1) optical properties on a large scale, 2) achievable 
levels of radioactive trace contaminations of  the PXE based scintillator, 
3) the performance of scintillator purification with a silica gel
column, and 4) of the liquid handling system, Module-0.
The purity levels required for detection of low-energy solar neutrinos,
in particular neutrinos from the $^7$Be electron capture,
are  at the $\mu$Bq/m$^3$ level corresponding to concentration of 
\uran\ and \thorium\ of $\sim10^{-16}$~g/g. For detailed specifications
of solar neutrino rates and background requirements, the
reader is referred to Refs.~\cite{BX-ST,BX-lowrad}.
As the $^{238}$U--progenies $^{222}$Rn, $^{210}$Pb, $^{210}$Bi and
$^{210}$Po are not necessarily in equilibrium with the progenitor activity,
we studied the contamination levels of \uran\ (as well as \thorium\ and several 
other isotopes) {\it directly} with neutron activation analysis (NAA) at levels of 
$10^{-16}$g/g and below. 
The short lived progenies are not accessible with NAA. Their contamination levels 
were studied in the CTF together with other backgrounds, as for example
the radiogenic produced $^{14}$C.
New techniques to remove radioactive isotopes 
by solid column purification were tested for the first time
in a ton scale experiment. Furthermore, the preparation, 
handling and purification of several tons of liquid scintillator
with Module-0 was the first test of a subsystem of the  
liquid handling system to be used in the scintillator 
operations in Borexino.

\subsection{Scintillator preparation}

About five tons of PXE solvent was procured 
from Koch Special Chemical Company,
Texas, USA.  To ensure a controlled procedure, 
the solvent was loaded by us at the company 
site in three specially modified and cleaned stainless steel
transport containers and shipped to the Gran Sasso underground laboratories. 

The PXE solvent was transferred from the 
transport containers into Module-0, a liquid handling and
purification system, specially built for the PXE
test.  It can be used for volumetric loading and 
unloading of liquid scintillator to/from the CTF 
Inner Vessel (IV), for fluor mixing, purification of liquid 
scintillator with a silica gel column, 
Rn degassing of liquid scintillator by nitrogen sparging,
and spray degassing as well as for   
water extraction.

Module-0 consists of a high and low pressure manifold system
which are connected by  pumps to build up the pressure difference 
of typically 2~atm. The manifolds are connected to tanks, columns, filters 
etc. to allow a variety of flow paths and operations.
The system includes two 7~m$^3$ electropolished pressure tanks (EP1/2),
two 1~m$^3$ process tanks (BT1/2) equipped with nitrogen 
spargers at the bottom and spray nozzles at the top 
(for a turbulent injection of liquid together with 
nitrogen gas),  pumps, 
flow meters, Millipore filters (0.5, 0.1 and 0.05\,micron) 
and one 70\,l high pressure column purification unit. All  
tanks are connected to a high purity nitrogen gas system to 
provide a nitrogen blanket at typically 30~mbar overpressure.
The complete system has been designed
according to ultra-high-vacuum standards to avoid contaminations
from the environmental air, in particular of $^{222}$Rn and $^{85}$Kr.
Only stainless steel and Teflon are in contact with the liquids. Metal
surfaces are electro polished and welds carried out with thorium free 
welding rods. Module-0 has been constructed in a class 1000 clean room. 
After completion,
all surfaces exposed to scintillator were cleaned following a detailed 
procedure to remove surface contaminations
as radon progenies $^{210}$Pb, $^{210}$Bi and $^{210}$Po. 
Details about the up-graded system are given in \cite{Mod0}.

The fluors (p-Terphenyl and bis-MSB, both from Sigma-Aldrich Co., 
scintillation grade) 
were sieved and then added without further purification 
to EP1 through a glove box connected to 
the top inlet flange which was flushed with nitrogen while the PXE solvent
was agitated with a nitrogen flux from the bottom inlet to facilitate
dissolving p-Tp. To accelerate the solvation of p-Tp,  the EP1 tank was 
heated to 38~$^o$C for four days under continuous nitrogen agitation.
The final scintillator mixture had a concentration of 
2.0\,g/l p-Tp and 20\,mg/l bis-MSB.

\subsection{Purification with silica gel}
\label{scint-purification}
One objective of the CTF test was to study the performance of 
scintillator purification with a solid silica gel column.
Preceding laboratory tests of the column purification with PXE/p-Tp with and 
without the addition of radio tracer showed a 
clear reduction of metal impurities ~\cite{thomas-dr}.   
The silica gel used during the CTF test has been 
radio assayed with HP germanium spectrometry and by direct measurements 
of the emanated radon with proportional counters. 
The results are listed in Table\,\ref{t:sigel-radon}. 
Despite the high bulk impurities of the 
Merck silica gel, we did not observe any measurable carry over 
into the scintillator apart from $^{222}$Rn. For further use in 
Borexino  we have found silica gel material with 
improved radio purity, in particular a radon emanation rate
of $0.13 \pm 0.07$~mBq/kg.

\begin{figure}
\epsfig{width=15cm,file=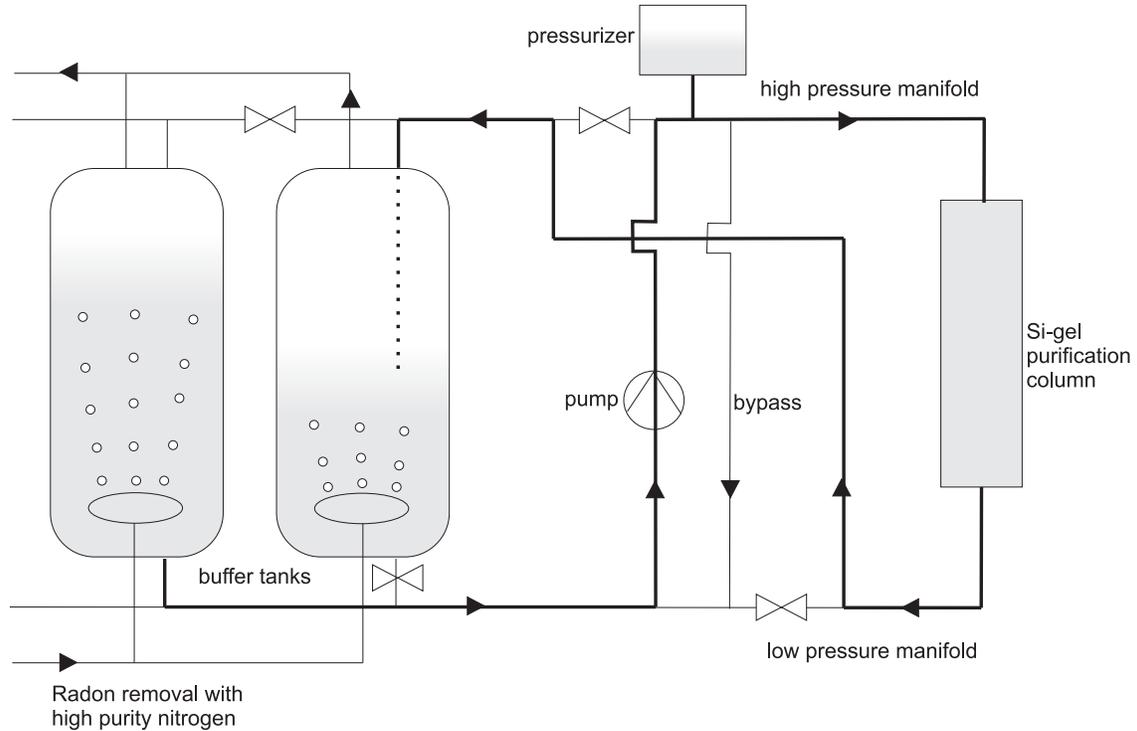}
\caption{\label{f:column}Flow path of PXE scintillator during column purification with 
subsequent nitrogen stripping in Module-0.}
\end{figure}

The scintillator components had been mixed as received from the supplier;
no special pretreatment had been given to the fluor and the wavelength shifter.
The ready mixed PXE scintillator was then passed through a column filled 
with silica gel (Merck, Silica Gel 60, 25-70 mesh ASTM) at a flow rate of 
typically 100~l/hour.  
Radon was removed by purging the scintillator after passing the column
as displayed schematically in Figure\,\ref{f:column}.
Samples for neutron activation analysis were taken prior and after all major 
steps of the scintillator preparation and purification:
{\bf Sample (1)} was collected from Module-0 after addition of p-Tp and 
bis-MSB to the PXE solvent (without purification).
{\bf Sample (2)} was collected during the filling of the IV. The complete scintillator  had passed once a 
column of about 80 cm height (15~kg) which had been exchanged against fresh
silica gel after the first two tons had run through 
the column, and once with a height of about 50 cm, i.e. in total two times.
{\bf Sample (3)} has been taken after the scintillator was circulated from the IV through the column
(re-filled to a height of 80 cm) and back into the IV. The flow rate 
was adjusted such that one cycle (5000 l) took two days. The scintillator
was circulated for four days, i.e. two cycles.
Subsequently the scintillator was 
unloaded from the IV  into the EP1-tank.
{\bf Sample (4)} was taken after completion of a water extraction 
in the EP1 tank. 
{\bf Sample (5)} was collected in Module-0 after circulating the scintillator
from EP1 through the column back to EP1. The column packing had not been 
exchanged (same as in (E)). This final operation lasted for eight days
with a flow rate of 3600 l/day, i.e. about 6 cycles.

\subsection{Radio assay with neutron activation analysis}

Combining neutron activation and low level counting methods, 
we developed a novel analytical method with sensitivities of $10^{-16}$g/g and below for
$^{238}$U and $^{232}$Th in liquid scintillators.
For this purpose about 250\,g samples of liquid scintillator were irradiated 
at a neutron flux of $\rm (10^{12} - 10^{13})\,s^{-1}cm^{-2}$ up to 100 hours
at the research reactor in Garching. The
long lived primordial radio nuclides are transformed 
into short-lived radio nuclides
(e.g. $\rm ^{238}U \to ^{239}Np$, \thorium $\to ^{233}$Pa), 
thus providing a higher specific activity with respect to 
their progenitors. After irradiation, liquid-liquid and ion exchange techniques were 
applied to separate the radio nuclides of interest from 
interfering activities. Coincidence counting methods \cite{isan}, 
e.g. ${\beta}$-${\gamma}$-conversion electron for 
$^{239}$Np and ${\beta}$-${\gamma}$ for $^{233}$Pa, 
further increase the sensitivity. Details can be found in 
Refs.~\cite{thomas,roger-como,BX-lowrad}.

The main results of the neutron activation analysis of the scintillator samples (1) to (5)
are summarized in Table~\ref{t:naaPXE1}~\cite{roger-dr}.
We report the concentrations of \uran, \thorium\ and other 
long-lived isotopes which contribute to the background in
Borexino.
In order to illustrate the performance 
of the silica gel column, also the concentrations for several
non-radioactive metal isotopes are given. 
An overall reduction between 1 and 3 orders of magnitude has been reached
for all elements where a positive value (above the detection limit) could be measured before the purification. 
Except for potassium, where the purity needed for Borexino is below the detection limit of the NAA,
the requirements for Borexino are met for all of the long lived radio nuclides.
New purity records in organic liquid 
scintillators have been achieved
for uranium at $\rm c(^{238}U)<1 \cdot 10^{-17}g/g$, and for thorium at
$\rm c(^{232}Th)<1.8 \cdot 10^{-16}g/g$.

\subsection{Measurements of PXE scintillator in the CTF}

The Borexino prototype detector CTF is a highly sensitive instrument
for the study of backgrounds in liquid scintillators at energies 
between a few tens of keV and a few MeV. 
It consists of a transparent 
nylon balloon with 2\,m diameter containing 4.2\,m$^3$ of liquid scintillator (Inner Vesssel, IV).
100 PMTs with light concentrators mounted on  a 7\,m diameter support
structure detect the scintillation signals (optical coverage 20\,\%).
The whole system is placed inside a cylindrical steel tank (11\,m in diameter, 10\,m height) 
that contains
1000\,tons of ultra-pure water in order to shield against external $\gamma$ rays from the 
PMTs and other
construction material as well as neutrons from the surrounding rock.
In the upgraded version of the CTF detector (after 1999), 
16 additional PMTs mounted on the floor of the water
tank detect the Cherenkov light created by muons transversing the water buffer 
and are used as a muon veto.
Details of the CTF detector and results of first measurements with pseudocumene
scintillator are given in Refs.~\cite{CTF-NIM,CTF-AP}. 

\subsubsection*{Sequence of measurements}
Direct counting after first loading of the PXE scintillator into the 
CTF in 1997 did not provide conclusive results because of 
photo-tube and electronics problems of the detector system that hindered data 
evaluation. 
After an upgrade of the CTF detector and the liquid handling system during 1998 and 1999, the 
same scintillator was filled again into the CTF in summer 2000, after passing once 
through a 40\,l silica gel column. 
The loading of the scintillator into the Inner Vessel 
was done in four batches of 1 ton each separated by short periods of 
data taking. 
Undisturbed data taking with 4.2 tons of PXE was going on from July 16 until September 5, 2000, 
in total 52 days. 
Afterwards, a series of calibration measurements
with a \radon\ point source were carried out where the source was 
moved inside the Inner Vessel to map the detector response and 
tune the position  reconstruction software.

\subsubsection*{Detector performances}
\begin{figure}
\epsfig{width=9cm,file=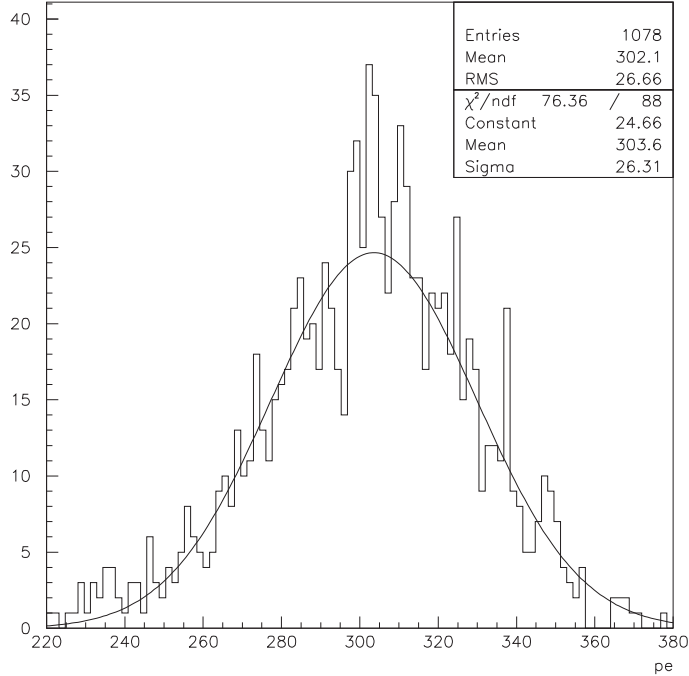}
\caption{\label{f:popeak}Pulse height distribution in units of photo-electrons (pe) 
of \polonium\ alpha decays in the CTF.}
\end{figure}

The pulse height\,-\,energy relation can be derived from the 
\polonium\ alpha peak (cf. Figure~\ref{f:popeak})
together with the measured quenching factors as given in Table\,\ref{t:quench}.
Due to its short half life  of 164\,$\mu$s, the \polonium\ decay can easily be tagged.
The \polonium\ stems mainly from \radon\ introduced during scintillator loading and therefore  
is  homogeneously distributed in the scintillator volume.
The alpha particle has an energy of 7.69\,MeV which in the scintillator is quenched to  
an equivalent beta energy of $(950\,\pm\,12)$\,keV. 
The measured peak position in the CTF corresponds 
to $ (304\,\pm\,3)$\,photo electrons leading to a yield of
$\rm (320\,\pm\,8)$~photo electrons per MeV.
The pulse height\,--\,energy relation is also a parameter of the 
\carbon\ fit (see below),
which is done in the energy range from 70 to 150\,keV, resulting in  
a somewhat higher value of 340~photo electrons per MeV. 
A more detailed analysis of the photo electron 
yield including a model with ionization quenching, 
has been presented in Ref.~\cite{nu-elm}. 
A value of $372 \pm 8$ photo electrons per MeV 
has been derived for the asymptotic yield, ie. for 
scintillation light created by  electrons 
with energies large compared to their rest mass.

The energy resolution derived from the width of the 
\polonium\ alpha peak corresponds to 
$\sigma(E)/E \simeq 8.7\,\%$, or $\sigma(E)/\sqrt{E}\simeq  2.6\, \rm keV^{1/2}$.
It is expected to scale with $\sqrt(E)$. 
A similar value of $\sigma(E)/\sqrt{E}\simeq  2.5\, \rm keV^{1/2}$   
is obtained from the spectral analysis of the 
\carbon\ spectrum.

\begin{figure}
\epsfig{width=14cm,file=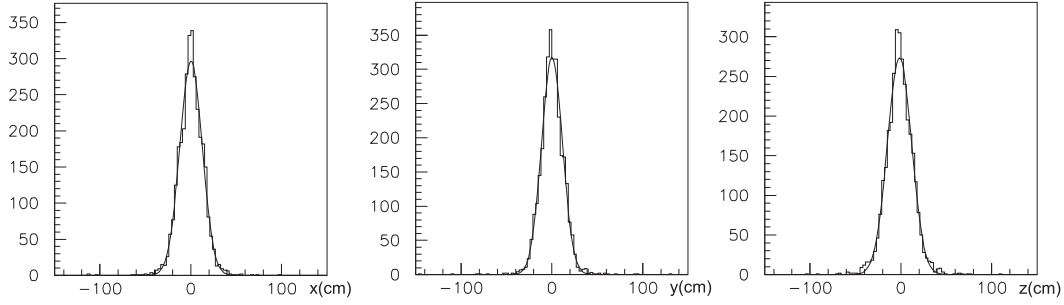}
\caption{\label{793resol}Reconstructed position (x,y,z) of $^{214}$Po events during a run 
with a $^{222}$Rn source at the detector's center.}
\end{figure}

The spatial resolution was studied with a localized \radon\ source at various 
positions inside the Inner Vessel \cite{marianne-dr}. 
Values of $\rm \sigma_{x,y}\simeq 12\,cm$, 
$\rm \sigma_{z}\simeq 13\,cm$ at 600\,keV were
obtained at the detector's center (cf. Figure~\ref{793resol}).
The spatial resolution degrades up to 15\,\%  close to the surface of the 
Inner Vessel.

\begin{figure}
\epsfig{width=7cm,file=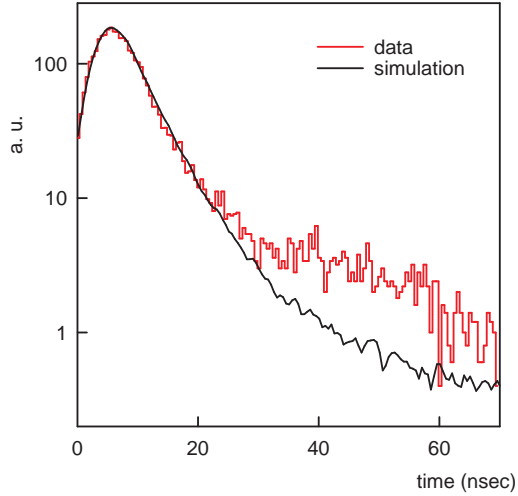}
\caption{\label{f:beta}Arrival time distribution of scintillation photons after 
excitation with $\beta / \gamma $   from  \bismuth\ decay  
at the detector's center. The red (grey) line corresponds to  CTF data and 
the black curve to MC simulation.}
\end{figure}

The photon arrival time distribution was studied with the source 
located at the center of the detector. Figure~\ref{f:beta}~ 
displays the time distribution for scintillation photons from the 
\bismuth\ beta decay. The data are compared with a Monte Carlo
simulation which includes the scintillation decay time 
from laboratory measurements given in Table~\ref{t:em}, 
elastic and inelastic scattering and the PMT time jitter. 
Monte Carlo and data show good agreement for arrival times up to 25\,ns. 
The data show a larger slow component than the MC simulation, which
could be due to reflected photons and/or PMT late pulses 
which are not fully accounted for in the simulation.

\begin{figure}
\epsfig{width=11cm,file=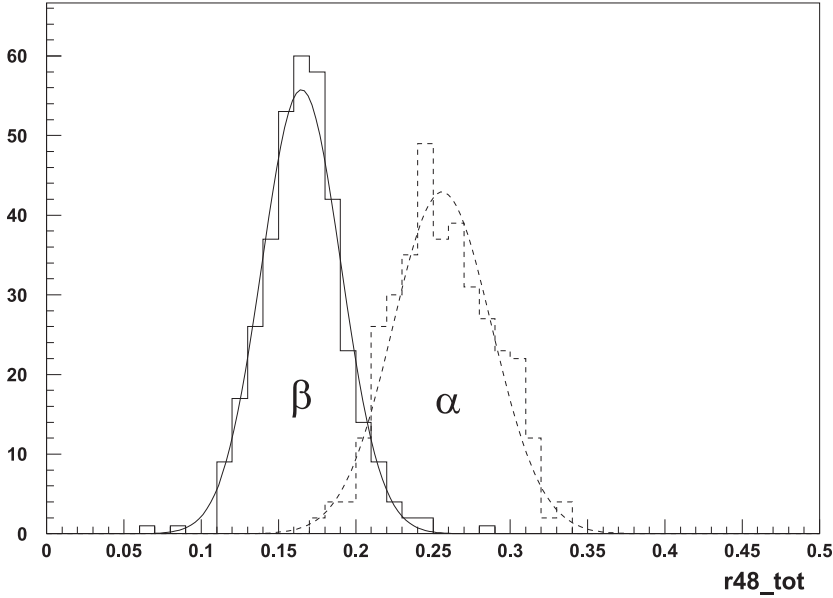}
\caption{\label{f:alphabeta}Tail-to-total-ratio r$_{48/tot}$  of \bismuth\ 
$\beta /\gamma$ and \polonium\ $\alpha$ 
events. Only $\beta /\gamma$ events with energy deposition 
between 0.6\,MeV and 1.3\,MeV  
are included.}
\end{figure}

To distinguish alpha from beta particles by pulse shape discrimination,
the analog sum of the charge signal from all PMTs is split into three
identical signals and fed into a charge sensitive ADC with different 
time delays. 
The total charge is derived from integration over the entire
charge signal 0 - 500\,ns, while the slow component is derived from 
the time windows 32 - 532\,ns and 48 - 548\,ns respectively.
The tail-to-total ratio r$_{32/tot}$ or r$_{48/tot}$ can  then  be used  
as discrimination parameter~\cite{alphabeta} as displayed in Figure~\ref{f:alphabeta}.  
The discrimination efficiency can be derived from 
a clean data sample of $\alpha$ and $\beta/\gamma$ events 
in the corresponding  energy range. 
Only the \bismuth($\beta$)-\polonium($\alpha$) decay 
sequence can unambiguously be identified as pure 
$\alpha$ or $\beta/\gamma$ events
due to the short half life of \polonium.
Only  \bismuth\ events in the same energy range as the 
\polonium\ events (0.6 - 1.3\,MeV) 
were considered.
Fixing the $\beta$ acceptance efficiency at 98~\% leads to an $\alpha$
identification efficiency of 84.6\,\%; 
applying a radial cut on the \bismuth\ and \polonium\ events of $r < 90$\,cm
leads to an increased $\alpha$ identification efficiency of 92.4\,\%.
This improvement is due to the fact that for events 
in the outer regions of the Inner Vessel the difference in the light path to 
different PMTs is higher, and therefore the fraction of light registered at late times
is higher.
Also other methods for $\alpha/\beta$ discrimination 
like the Gatti optimum filter method have 
been studied with the CTF \cite{maria-elena,tristan}, and 
will be implemented in Borexino.

\subsubsection*{Radiopurity analysis}
\begin{figure}
\epsfig{width=10cm,file=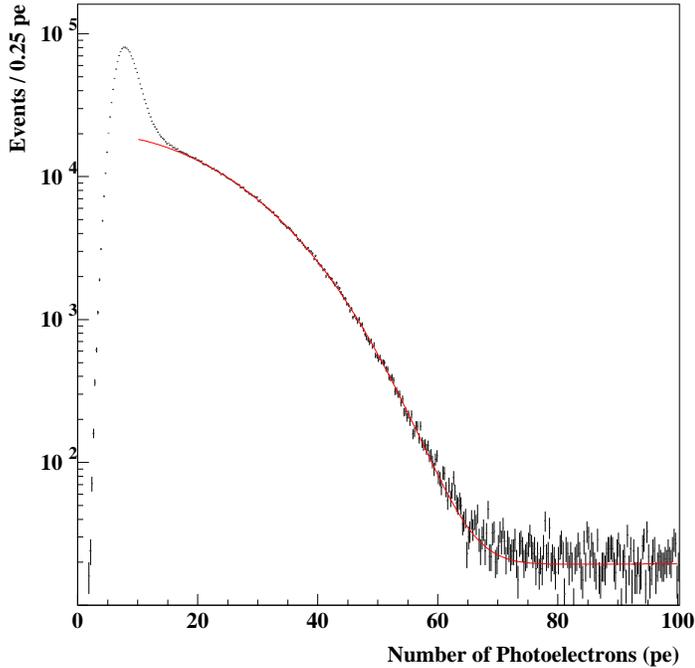}
\caption{\label{f:c14-ctf2}The low energy spectrum collected during a period of 6.4 days. 
The counting rate is dominated by the \carbon\ $\beta$--decay with  
$\rm ^{14}C/^{12}C =(9.1  \pm 0.3\,(syst.) \pm 0.3\,(stat.))
\cdot 10^{-18}$. 
The solid line shows the fit of the theoretical \carbon\ spectrum to the data.}
\end{figure}

The counting rate below 200\,keV is dominated by
the \carbon\ beta decay with an endpoint of 156\,keV.
The \carbon\ concentration in the scintillator was determined by fitting a
convolution of the theoretical beta spectrum and the energy dependent detector 
resolution function, plus a background contribution to the measured spectrum 
in the energy range from 70 to 150\,keV \cite{CTF-C14,elisa-dott}. 
The low energy spectrum together with the fit is displayed in 
Figure~\ref{f:c14-ctf2}.
At energies below 70\,keV there is a background contribution from 
Cherenkov events produced by $\gamma$'s in the shielding water.
From the fit we derive the \carbon\ activity which translates to a ratio of  
$\rm ^{14}C/^{12}C =(9.1  \pm 0.3 (\rm stat)\pm 0.3 (\rm syst))\times 10^{-18}$. 
The systematic error is dominated by the uncertainty of the total 
scintillator mass.

The \uran\  decay chain is assayed via the delayed coincidence of the \radon\ progenies 
 \bismuth\ ($\beta/ \gamma$) and \polonium\ ($\alpha$) with a half life of 164\,$\mu$s.
Assuming secular equilibrium of the decay chain -- which is likely to be broken due to the long-lived 
\radium\ and the gaseous \radon --
the measured \bismuth\--\polonium\ activity can be expressed as uranium-equivalent.
After loading the CTF  an initial \bismuth\--\polonium\ rate of  about
90 counts per day was observed which was clearly related to \radon\ introduced during the filling process, as it decayed away with the \radon\ half life of 3.8 days. 
The last two 1-week-periods  of data taking with a total counting
time of 7.32 days (period I) and 6.98\,days (period II) were used for the analysis, 
when the contribution from the initial \radon\ contamination was below 1 count per day.
The detector efficiency (due to the dead time connected with the read out of each event) 
was 91\,\% during that period. 
The \bismuth-\polonium\ events were selected by applying the following cuts:
no muon veto trigger (efficiency $>$ 99\,\%);
energy of the first event $>$ 300\,keV (efficiency 95\,\%);
energy of the following event between 0.6 and 1.25\,MeV (99\,\%);
coincidence time between 5\,$\mu$s and 800\,$\mu$s (94\%). 224 (199) 
candidate events survived the cuts for period I (II), corresponding to a \radon\ activity
of 102\,$\mu$Bq/m$^3$ (99\,$\mu$Bq/m$^3$).
This activity however was not homogeneously distributed within the 
scintillator volume, but 
most of the events were localized along 
the vertical symmetry axis of the detector. 
An artefact due to false reconstruction
could be excluded after calibration with point like sources which 
were located both on and off axis. Though the origin of these events
could not be fully resolved, it is excluded that they are
related to the intrinsic scintillator impurities.
To derive a number for the residual
\radon\ concentration homogeneously distributed in the scintillator, 
 a cylindrical cut around the vertical axis was applied. 
For $\rm R_{x,y}>0.6\,m$ we get 
a \radon\ activity of $\rm A(^{222}Rn) = (27 \pm 5) \mu Bq/m^3$ in period I
and $\rm A(^{222}Rn) = (23 \pm 5) \mu Bq/m^3 $ in period II,
or a \uran\ equivalent concentration of 
 $\rm c(^{238}U) = (2.3 \pm 0.4) \cdot 10^{-15}\,g/g \quad (I)$
and $\rm c(^{238}U) = (1.9 \pm 0.4) \cdot 10^{-15}\,g/g \quad (II)$\,.
Due to the above stated \radon\ problem this value must be considered as an
upper limit of the internal \uran\ contamination. 

Further radio-isotopes from the uranium decay chain of concern 
are the $^{210}$Pb progenies $^{210}$Bi and $^{210}$Po. Secular 
equilibrium typically is disturbed even within this sub-chain, 
as well as with respect to the progenitor $^{226}$Ra 
because of the characteristic lifetimes 
and the different chemistries involved. 
From a spectral analysis, about 100 to 200 decays per day with alpha-like pulse shapes and 
with energy depositions 
quenched to approximately 0.5~MeV are attributed
to  $^{210}$Po decays in the scintillator \cite{laura-phd}.

A limit for the intrinsic \thorium\ contamination can be derived  
via the delayed $\beta$-$\alpha$ coincidence of its progenies $^{212}$Bi-$^{212}$Po 
 with a half life of 299\,ns. 
It can be distinguished from
the \bismuth-\polonium\ coincidence by the higher energy of the 
$\alpha$ decay and the shorter coincidence time, 
though the latter has to be considered as a background.
Secular equilibrium in the \thorium\ chain is usually observed  
for $^{228}$Th and its progenies. 
The following cuts were applied to select the $^{212}$Bi-$^{212}$Po
events: no muon veto trigger;
energy of the first event $>$ 300\,keV (efficiency 85\,\%);
energy of the second event $>$ 800\,keV (100\,\%);
coincidence time between 50 and 1500\,ns (86\,\%);
due to a contamination at the bottom of the vessel only decays in the upper hemisphere of the detector were considered (50\,\%). 
The overall efficiency of the cuts for the selection of the $^{212}$Bi-$^{212}$Po events
is 37\,\%.
During the whole period of undisturbed data taking, in total 25.8\,days of live time,
12 candidates were detected, leading to a \thorium\ equivalent concentration of
$ \rm c(^{232}Th) =  (1.3 \pm 0.4) \cdot 10^{-15}  g/g\,$.

\begin{figure}
\epsfig{width=14cm,file=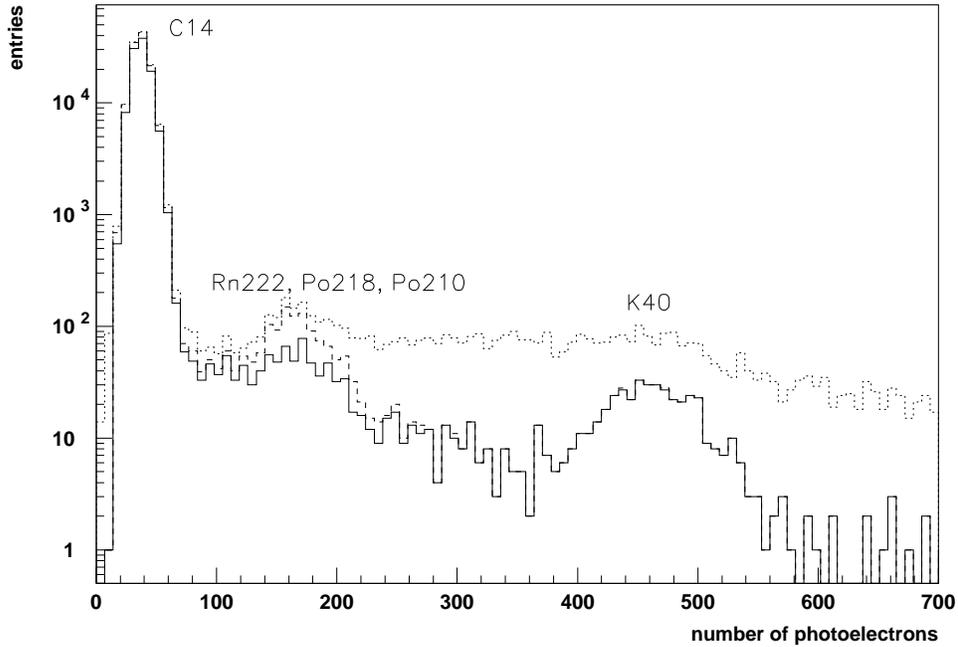}
\caption{\label{internalback}PXE data collected in a period of 7.3 days. 
Only single events with $r<90$\,cm
are shown (dotted line). In the next step, all muon events are 
removed (dashed line). In the final step 90\,\% of the alpha events 
are subtracted via pulse shape analysis (solid line). }
\end{figure}

An estimation of the remaining background for the neutrino 
detection in Borexino
is provided by the subtraction of all known contributions from the 
background spectrum measured in the CTF.
In order to reduce the contribution from the external background, a radial cut ($r<0.9$\,m)
has been applied. Also, all correlated events have been tagged and 
discriminated.
In the neutrino window (NW, $\rm 250\,keV <  E \rm < 800\, keV$), the counting rate then is
120 events per day and ton. 
As next step, all muon induced events have been subtracted (74 events per day and ton).
In this inner region of the detector, alphas can be discriminated via their pulse shape
with an efficiency of $\sim 90$\,\%, leading to 48 events per day and ton.
The resulting background spectrum is shown in Figure~\ref{internalback}.
Clearly visible are the contributions from \carbon\, the remaining 
alphas from the 
\uran\ and \thorium\ chain, 
and a \kalium\ $\gamma$ peak, which could be identified as external 
background originating from the  Vectran strings 
that hold the Inner Vessel in place (a dedicated measurement of the strings showed a concentration of 45 
ppm K).

\subsubsection*{Summary of Results}
The large scale test of a scintillator based on PXE in the CTF showed 
that its optical properties and the detector performance are comparable to 
a pseudocumene based scintillator.
The purity levels required for the detection of low energy solar neutrinos can be 
achieved by purification of the scintillator with a silica gel column.
Though the very low \uran\ and \thorium\ concentrations  
measured by NAA ($< 1\cdot 10^{-17}$\,g/g and $< 2\cdot 10^{-16}$\,g/g ) could 
not be confirmed by the CTF measurements, 
the remaining activity levels were close to the detector's sensitivity limit and largely
influenced by systematics e.g. due to external background and surface contamination.

\section{Outlook}
PXE based scintillator has been investigated for application in Borexino as  
detector medium for low energy solar neutrino spectroscopy. 
Key questions under 
study were the optical properties, radio purity and new techniques 
of purification
by solid column extraction. Both laboratory measurements and the large scale 
operations in the CTF showed that PXE scintillator as target and water 
as buffer medium 
is a viable solution for Borexino and it was rated as a backup solution. The 
default configuration in the Borexino design is based on pseudocumene as 
scintillator as well as buffer medium. 

Because of its superior 
safety characteristics and the better self-shielding due to its higher density, 
future applications of PXE in underground experiments are conceivable, 
as for example the search 
for non-vanishing value of $\theta_{13}$ with reactor neutrinos, 
the search for neutrinoless double beta decay with metal loaded scintillators,
the study of neutrinos from the interior of the Earth and from Supernovae,
as well as the study of proton decay via the K-meson decay channel.

\section{Acknowledgements}
F.X. Hartmann would like to thank the extra-ordinary services of the 
Koch Speciality Chemical Company and the ALCOA Company as regards the 
solid column design.

\newpage

\begin{table}
\caption{\label{t:pxe-chem-phys}Physical and chemical data of PXE. $^1$: Pensky-Martens closed cup.}
\begin{tabular}{|ll|}
\hline 
cas no.                     & 6196-95-8 \\
formula                     & $\rm C_{16} H_{18}$ \\
molecular weight            & 210.2 g/mol \\
conc. of ortho isomer       & 99 \% \\
range of density (15$^o$C)  & 0.980 - 1.000 g/cm$^3$\\
typical density  (15$^o$C)  & 0.988 g/cm$^3$ \\
vapor pressure (20$^o$C)    & $<0.00014$ hPa \\
vapor pressure (80$^o$C)    & 0.13 hPa \\
boiling point               & 295$^o$C  \\
flash point$^1$             & 145$^o$C  \\
auto-ignition               & 450$^o$C  \\ 
viscosity (40$^o$C)         & 5.2 cSt   \\
solubility in water (20$^o$C) & 0.01 g/l \\
acid no.                    & $<0.005$ mg KOH/g \\ 
refraction index            & 1.565 \\
\hline
\end{tabular}
\end{table}

\begin{table}
\caption{\label{t:em}Parameters of the time profile for photon emission  
after excitation with alpha and beta particles. 
a: PXE/p-Tp(2.0 g/l)/bis-MSB(20~mg/l)
b: PXE/p-Tp(3.0 g/l)/bis-MSB(20~mg/l).}
\begin{tabular}{|c|cccc|cccc|}
\hline 
\bf excitation &$\tau_1$ & $\tau_2$ &$\tau_3$ &$\tau_4$ & $q_1$ & $q_2$ & $q_3$ & $q_4$ \\
\hline
beta$^{(a)}$ & 3.8 & 15.9 & 63.7 & 243.0 & 0.832 & 0.114 & 0.041 & 0.013\\
\hline 
beta$^{(b)}$ & 3.1 & 12.4 & 57.1 & 185.0 & 0.788 & 0.117 & 0.070 & 0.025\\
\hline 
alpha$^{(b)}$ & 3.1 & 13.4 & 56.2 & 231.6 & 0.588 & 0.180 & 0.157 & 0.075 \\
\hline
\end{tabular}
\end{table}

\clearpage
\newpage

\begin{table}
\caption{\label{t:quench}Quenching factors for alpha particles with 
different energies for PXE/p-Tp(2 g/l)/bis-MSB(20~mg/l).}
\begin{tabular}{|c|c|c|c|} \hline
Element & $\alpha$-energy [MeV]& measured energy [keV]& quenching factor \\ 
\hline 
$^{210}$Po & 5.30 & $(490\pm 10)$ & $(10.8\pm 0.2)$ \\
$^{222}$Rn & 5.49 & $(534\pm 10)$ & $(10.3\pm 0.2)$ \\ 
$^{218}$Po & 6.00 & $(624\pm 10)$ & $(9.6\pm 0.2)$ \\ 
$^{214}$Po & 7.69 & $(950\pm 12)$ & $(8.1\pm 0.1)$ \\ 
 \hline 
\end{tabular} 
\end{table}

\begin{table}
\caption{\label{t:sigel-radon}Trace contaminations determined by HP-Ge spectrometry and 
radon emanation ($^\ast$) 
of Silica Gel 60 from Merck.}
\begin{tabular}{|c|c|}
\hline 
\radium\    & $2.28 \pm 0.11$ Bq/kg \\
\radon\ $^\ast$   & $200 \pm 10$  mBq/kg  \\
\uran\      & $1.6 \pm 0.3$ Bq/kg   \\
$^{210}$Pb  & $1.6 \pm 0.3$ Bq/kg   \\
$^{228}$Th  & $1.4 \pm 0.1$ Bq/kg   \\
$^{228}$Ra  & $1.4 \pm 0.1$ Bq/kg   \\
$^{40}$K    & $1.9 \pm 0.4$ Bq/kg   \\
$^{137}$Cs  & $<0.040 $     Bq/kg   \\
\hline
\end{tabular}
\end{table}

\tiny
\begin{table}
\caption{\label{t:naaPXE1}Concentration of trace elements in the PXE scintillator.
The samples  were taken during the scintillator 
preparation and purification 
processes (cf. Sec.~\ref{scint-purification}). The upper part lists elements with long lived radioactive isotopes which are a potential background in Borexino, the lower part lists some
non-radioactive elements for illustration of the cleaning performance.
ND: no data, limits: 90\% CL, errors: $1\sigma$.
}
\hskip -2cm
\begin{tabular}{|c|c|c|c|c|c|}
\hline
{\bf Ele-}& \multicolumn{5}{c|}{\bf concentration in g/g}  \\ \cline{2-6}
\bf ment &\bf  1 
&  \bf 2 & \bf 3 &\bf 4 & \bf 5 \\
\hline
{\bf Th} &  $< 2 \cdot 10^{-14}$ &  $ (5.0\pm 2.5) \cdot 10^{-15}$ &
$<3 \cdot 10 ^{-15}$ & $ (2.5\pm 0.7) \cdot 10^{-16}$ &  $<1.8  \cdot 10^{-16}$\\
{\bf U}  & $(4.0\pm 2.0) \cdot 10^{-14}$ & $(1.4\pm 0.7) \cdot 10^{-15} $ &
$<6 \cdot 10 ^{-16}$ & $< 2.5 \cdot 10^{-16}$ &  $ <1 \cdot 10^{-17}$\\
Cd   & ND & $< 2 \cdot 10^{-12}$ & ND & $< 5.4 \cdot 10^{-14}$ & $< 8.3 \cdot 10^{-15}$ \\

In  & $< 3 \cdot 10^{-13}$ &  $(5.0\pm 2.5) \cdot 10 ^{-12}$ &$< 4.8 \cdot 10 ^{-13}$ &
$< 2.5 \cdot 10 ^{-14}$ & $< 1.2 \cdot 10 ^{-13}$ \\
K   &  $< 8 \cdot 10 ^{-12}$  & $<5\cdot 10^{-11}$ & 
$< 1.3 \cdot 10 ^{-11}$&$< 2 \cdot 10 ^{-12}$ &$< 6.1 \cdot 10 ^{-12}$   \\
La  &$< 3 \cdot 10^{-14}$ &$< 2 \cdot 10 ^{-14}$     & $ <6.4 \cdot 10^{-15}$ & $< 9.4 \cdot 10 ^{-16}$&
$< 4.3 \cdot 10 ^{-16}$   \\
Lu  & $<2 \cdot 10 ^{-15}$ & $< 7 \cdot 10 ^{-16}$&$< 1.4 \cdot 10 ^{-15}$
& $< 4.0 \cdot 10 ^{-16}$&$< 3.8 \cdot 10 ^{-16}$     \\
Rb  & $< 3 \cdot 10^{-13}$ & $< 8 \cdot 10^{-12}$ 
& $< 2.8 \cdot 10 ^{-13}$ & $< 1.1 \cdot 10 ^{-13}$ & $< 1.2 \cdot 10 ^{-13}$   \\
\hline
Ag & $<1 \cdot 10 ^{-12}$ & $<2 \cdot 10 ^{-12}$ & $(7.9\pm 3.0) \cdot 10 ^{-14}$ &  
$(1.1\pm 0.5) \cdot 10 ^{-13}$ & $(2.3\pm 0.4) \cdot 10 ^{-13}$\\
Au &  $(2.0\pm 1.0) \cdot 10 ^{-14}$ & $(2\pm 1) \cdot 10 ^{-15}$& ND & $(1.8\pm 0.9) \cdot 10 ^{-16}$ 
& $< 3.8 \cdot 10 ^{-17}$\\
Cr & ND   & $(3\pm 1.5) \cdot 10^{-11}$ &$< 1 \cdot 10^{-13}$     & $ (1.9\pm 0.9) \cdot 10^{-13}$ & 
$ (2.3\pm 0.4) \cdot 10^{-13}$ \\
Fe & $< 2 \cdot 10^{-10}$ & $< 2 \cdot 10^{-10}$ & ND & $< 2.5 \cdot 10^{-12}$ & $(3.7\pm 0.6) \cdot 10^{-11}$  \\
Sb & $ (3.0\pm 1.5) \cdot 10^{-13}$ & $(3.0\pm 1.5) \cdot 10^{-13}$ & ND
& $(3.0\pm 1.5)\cdot 10 ^{-14}$ & $(8.7\pm 1.3) \cdot 10 ^{-15}$   \\
W & ND & $(1.0\pm 0.5) \cdot 10^{-13}$ &  
ND & $(1.4\pm 0.9) \cdot 10^{-14}$ &  $<4.8 \cdot 10^{-15}$\\
Zn  & ND & $(7.0 \pm 3.5) \cdot 10^{-11}$ & ND &  
$(3.1\pm 0.5) \cdot 10^{-13}$ &  $(1.0\pm 0.2) \cdot 10^{-12}$\\
\hline
\end{tabular}
\end{table}


\begin{thebibliography}{00}


\bibitem{BX-ST}
G. Alimonti {\it et al.} (Borexino Collaboration),
Astropart. Phys. 16 (2002) 205-234 

\bibitem{CTF-NIM} 
G. Alimonti {\it et al.} (Borexino Collaboration),
Nucl. Inst. Meth. {\bf A} 406 (1998) 411-426  


\bibitem{CTF-C14} 
G. Alimonti {\it et al.} (Borexino Collaboration),
Phys. Lett. {\bf B} 422 (1998) 349-358 

\bibitem{CTF-AP}
G. Alimonti {\it et al.} (Borexino Collaboration),
Astropart. Phys. 8 (1998) 141-157 



\bibitem{KL}
K. Eguchi {\it et al.} (KamLAND Collaboration), 
Phys. Rev. Lett. 90  (2003) 021802


\bibitem{CFR}See for example: United States Code of Federal Regulations,
Title 49, Vol. 2, part 173 

\bibitem{Majewski} S. Majewski et al., NIM {\bf A} 414 (1998)  289-298 

\bibitem{doublechooz}F. Ardellier at al. (Double Chooz collaboration), hep-ex0606025

\bibitem{Mod0}
F.X. Hartmann, Proc. of the 4th Int. Neutrino Conference, 
Heidelberg (1997), ed by W. Hampel, MPIK Heidelberg, p. 202.\\  
L. Niedermeier {\it et al.}, NIM {\bf A} 568 (2006) 915-922


\bibitem{el-decay} H. O. Back {\it et. al.} (Borexino Collaboration),
Phys. Lett. {\bf B} 525 (2002)  29-40 

\bibitem{nucl-decay} H. O. Back {\it et. al.} (Borexino Collaboration),
Phys. Lett. {\bf B} 563 (2003) 23-34 

\bibitem{nu-elm} H. O. Back {\it et. al.} (Borexino Collaboration),
Phys. Lett. {\bf B} 563 (2003)  35-47 

\bibitem{pauli-excl} H. O. Back {\it et. al.} (Borexino Collaboration),
Eur. Phys. J. C  37 (2004) 421-431  

\bibitem{tesilombardi} P. Lombardi, Diploma thesis, Universita di Milano (1996)

\bibitem{motta} D. Motta, PhD thesis, MPIK Heidelberg (2004)

\bibitem{neff-da}M. Neff, Diploma thesis, 
Technische Universit\"at M\"unchen (1996)

\bibitem{llbf} C. Buck, F.X. Hartmann, D. Motta, S. Sch\"onert, U. Schwan, 
Nucl. Phys.{\bf B} (Proc. Suppl.) 118 (2003)  451 

\bibitem{buck} C. Buck, PhD thesis, MPIK Heidelberg (2004)




\bibitem{BX-lowrad} C. Arpesella {\it et al.} (Borexino Collaboration),
Astropart. Phys. 18 (2003) 1-25  


\bibitem{thomas-dr}T. Goldbrunner, PhD thesis, 
Technische Universit\"at M\"unchen (1997)



\bibitem{isan}R.S. Raghavan et al., AT $\&$ T Bell Laboratories, Tech.
Memorandum, {\bf 11121-921015-37} (1992)

\bibitem{thomas}T. Goldbrunner {\it et al.}, Nucl. Phys. B (Proc. Suppl.) 
{\bf  61} (1998) 176 

\bibitem{roger-como}R.~v.~Hentig et al., Nucl. Phys. B (Proc. Suppl.) {\bf 78}
    (1999)  115-119
	 
\bibitem{roger-dr}R.~v.~Hentig, PhD thesis, 
Technische Universit\"at M\"unchen (1999)

\bibitem{marianne-dr}M. G\"oger-Neff, PhD thesis, 
Technische Universit\"at M\"unchen (2001)

\bibitem{alphabeta}G. Ranucci et al.,  NIM {\bf A} 412 (1998) 374 

\bibitem{maria-elena} M. Monzani, Diploma thesis, Universita di Milano (2001)

\bibitem{tristan} T. Beau, PhD thesis, Universite Paris 7 (2002)

\bibitem{elisa-dott}
E. Resconi, PhD thesis, Universita di Genova (2001)

\bibitem{laura-phd}
L. Cadonati, PhD thesis, Princeton University (2000)


		
\end{thebibliography}
\end{document}